%% file: paper.tex
\documentclass[sigconf, authorversion=true]{acmart}
\renewcommand\footnotetextcopyrightpermission[1]{} 
\def\BibTeX{{\rm B\kern-.05em{\sc i\kern-.025em b}\kern-.08emT\kern-.1667em\lower.7ex\hbox{E}\kern-.125emX}}
    
\usepackage{enumitem}
\usepackage{tikz}
\usepackage{booktabs} 
\usepackage{listing}
\usepackage{listings}
\usepackage{subfloat}
\usepackage{subfig}
\usepackage{hyperref}
\usepackage{breakurl}
\usepackage{enumitem}
\usepackage{algorithm}
\usepackage{algpseudocode}
\usepackage{booktabs}
\usepackage[most]{tcolorbox}
\usepackage{makecell}
\usepackage{multirow}
\usepackage{xspace}
\usepackage{pifont}
\usepackage{balance}
\usepackage{xcolor}
\usepackage{graphics}
\usepackage{pdfpages}
\usepackage[]{soul}
\usepackage{textcomp} 
\usepackage[htt]{hyphenat} 
\setcopyright{none}

\begin{document}

%
\title[Insecure Until Proven Updated]{Insecure Until Proven Updated:\\Analyzing AMD SEV's Remote Attestation}

%
\author{Robert Buhren}
\email{robert.buhren@sect.tu-berlin.de}
\affiliation{%
  \institution{Technische Universität Berlin\\Security in Telecommunications}
}
\author{Christian Werling}
\email{christian.werling@student.hpi.de} 
\affiliation{%
  \institution{Hasso Plattner Institute, Potsdam}
}
\author{Jean-Pierre Seifert}
\email{jpseifert@sect.tu-berlin.de}
\affiliation{%
  \institution{Technische Universität Berlin\\Security in Telecommunications}
}

%

%
\begin{abstract}
Cloud computing is one of the most prominent technologies to host Internet services that unfortunately leads to an increased risk of data theft.
Customers of cloud services have to trust the cloud providers, as they control the building blocks that form the cloud.
This includes the hypervisor enabling the sharing of a single hardware platform among multiple tenants.
Executing in a higher-privileged CPU mode, the hypervisor has direct access to the memory of virtual machines.
While data at rest can be protected using well-known disk encryption methods, data residing in main memory is still threatened by a potentially malicious cloud provider.

AMD Secure Encrypted Virtualization (SEV) claims a new level of protection in such cloud scenarios.
AMD SEV encrypts the main memory of virtual machines with VM-specific keys, thereby denying the higher-privileged hypervisor access to a guest's memory.
To enable the cloud customer to verify the correct deployment of his virtual machine, SEV additionally introduces a remote attestation protocol.
This protocol is a crucial component of the SEV technology that can prove that SEV protection is in place and that the virtual machine was not subject to manipulation.

This paper analyzes the firmware components that implement the SEV remote attestation protocol on the current AMD Epyc Naples CPU series.
We demonstrate that it is possible to extract {\em critical} CPU-specific keys that are fundamental for the security of the remote attestation protocol.

Building on the extracted keys, we propose attacks that allow a malicious cloud provider a complete circumvention of the SEV protection mechanisms.
Although the underlying firmware issues were already fixed by AMD, we show that the current series of AMD Epyc CPUs, i.e., the Naples series, does not prevent the installation of previous firmware versions.
We show that the severity of our proposed attacks is very high as no purely software-based mitigations are possible.
This effectively renders the SEV technology on current AMD Epyc CPUs useless when confronted with an untrusted cloud provider.

To overcome these issues, we also propose robust changes to the SEV design that allow future generations of the SEV technology to mitigate the proposed attacks.
\end{abstract}

%
%


\keywords{virtualization, Secure Encrypted Virtualization, cloud computing, shielding systems, SEV, remote attestation}

%

%
\maketitle

\input{src/1_introduction}
\input{src/2_background}

\input{src/3_firmware_analysis}
\input{src/4_attack_motivation}
\input{src/5_attack}

\input{src/6_discussion}
\input{src/7_related_work}
\input{src/8_conclusion}

%
\begin{acks}
This work was supported by the Federal Ministry of Education and Research of Germany in the framework of Software Campus 2.0 project no. FKZ 01IS17052.
Opinions, views, and conclusions are those of the authors and do not reflect the views of anyone else.
We would like to thank the following persons for their help during the work on this paper: Peter Stuge, Stephan Bauroth, Nils Wisiol, Elham Amini and Heiko Lohrke.
\end{acks}

%
\bibliographystyle{ACM-Reference-Format}
\bibliography{paper}

%

\end{document}

%% file: src/1_introduction.tex
\section{Introduction}

Cloud computing is one of the core foundations of today's Internet landscape. 
The manifold advantages such as on-demand resource allocation or high availability of services have lead to a wide usage of this technology.
However, outsourcing the processing of enterprise data comes at a risk.
The technical infrastructure that forms the cloud is owned by the cloud provider and thus under his full control.
This includes the server hardware, as well as the software components that allow the co-location of multiple virtual machines on a single host.

Therefore security concerns impede the deployment of confidential data and applications in cloud scenarios \cite{Hashizume2013, Kandias2013}.
The potential threats range from misconfiguration of software components over cloud provider admin access to foreign government access \cite{TheVerge2018}.

To counter these threats, the research community, as well as industry, proposed new approaches to allow secure cloud computing when confronted with an untrusted cloud provider \cite{cloudvisor2011, hsvm2011, flicker2008, hypercoffer2013, hyperwall2012, fidelius2018}.
Most prominent is the recently released \textit{Secure Encrypted Virtualization} technology by AMD \cite{AMD2016}.
SEV's goals are twofold: 

\begin{enumerate}[label=(\alph*)]
	\item Prove the correct deployment of virtual machines.
	\item Offer virtual machine protection at runtime.
\end{enumerate}

To achieve (a), a dedicated co-processor, the \emph{Platform Security Processor} (PSP), creates a cryptographic hash of the components that form the initial virtual machine state, similar to the remote attestation feature of a TPM.

\begin{quote}
	``With this attestation, a guest owner can ensure that the hypervisor did not interfere with the initialization of SEV before transmitting confidential information to the guest.''~\cite{AMD2016}
\end{quote}

(b) is achieved by using memory encryption with virtual machine-specific encryption keys.
These keys are generated upon creation of a virtual machine and are stored inside the PSP that uses its own private memory.
This memory is not accessible from the main processor, which is executing software components controlled by the cloud provider such as the hypervisor.
A hardware memory encryption unit provides transparent encryption and decryption using the virtual machine-specific key when the respective virtual machine is scheduled.

\begin{quote}
	``Even though the hypervisor level is traditionally 'more privileged' than the guest level, SEV separates these levels through cryptographic isolation.''~\cite{AMD2016}
\end{quote}

Furthermore, to allow the authentication of an SEV platform, each SEV-enabled platform contains a key pair that is unique to this platform.	 
The public key is signed by AMD and, given a platform-specific ID, a guest owner can obtain this key from an AMD key server.
This key is the basis for a remote attestation protocol that enables the user to verify the correct deployment of her VM, including that SEV protection is in place.
Only then she will inject a guest secret, e.g., a disk encryption key, into the guest VM using a secure channel between the PSP and the guest owner.
This remote attestation feature is a key component when using the infrastructure of an untrusted cloud provider.
It provides cryptographic proof that the cloud provider uses an authentic AMD platform with SEV enabled and deployed the virtual machine according to the owner's configuration.

Previous research focused on the security of SEV-protected virtual machines at runtime \cite{Hetzelt:2017, du2017secure, DBLP:journals/corr/abs-1805-09604} while the remote attestation feature of SEV has not been subject to a comprehensive analysis yet.
To overcome this gap, we analyzed the firmware components that implement the remote attestation feature\footnote{To facilitate further research we published a PSP firmware analysis tool. The tool can be found at~\cite{psptoolGithub}.}.
We demonstrate that it is possible to extract the platform-specific private key, upon which the security of the remote attestation protocol depends.
This enables the untrusted cloud provider to violate the security goals of the SEV technology.
Specifically, it enables the cloud provider to forge the presence of SEV altogether, intercept the communication between guest owner and SEV firmware and decrypt guest memory.

We show that it is, to the best of our knowledge, impossible to provide a purely software-based fix for these issues.
This questions the security capabilities of the SEV feature for the complete AMD Epyc Naples processor line.
More so, it actively puts SEV customers' data at risk since SEV might attract confidential data that was previously not hosted in the cloud. 

Furthermore, we propose a new design for the remote attestation protocol. 
While the current SEV design cannot cope with firmware security issues, the proposed design allows to revert to a trusted state in case of previous firmware issues.
Our proposed design enables SEV customers to trust the remote platform even in case the platform-specific private key was leaked. \\

\noindent\textbf{\textit{Our contributions are:}}\quad
\begin{enumerate}
    \item We conduct a comprehensive security analysis of the firmware components that implement SEV's remote attestation protocol.
      Our analysis reveals issues that allow to extract private keys that are fundamental to the security of the SEV technology.
      More so, the current AMD Epyc CPUs allow to install arbitrary signed firmware versions, hence extraction is possible even on systems that use patched firmware versions.
    \item We propose attacks based on our findings that allow to fake the presence of SEV or extract encrypted VM memory in plaintext.
    \item We show a severe design issue of the protocol, rendering it useless in case of common firmware issues. The severity is amplified as the issue allows to mount attacks targeting platforms with no security issues present: \emph{Possession of an extracted key is sufficient to mount the attacks, regardless of whether the key belongs to the attacked platform or not}.
    \item Lastly, we propose robust design changes to the SEV technology that allow future generations of SEV to mitigate the impact of firmware security issues. 
\end{enumerate}

\noindent\textbf{\textit{Ethical considerations:}}\quad
We informed AMD of the firmware issues found during our analysis.
AMD confirmed the presence of the firmware issues and published security updates to the PSP firmware.
Similar issues had been previously reported by other security researchers \cite{CTSLABS2018}.
Furthermore, we reported the extracted keys to AMD to allow a revocation of the corresponding certificates.
Although security fixes are already provided by AMD, our findings show that software-based mitigations are not sufficient.

\begin{quote}
We therefore refrain from publishing any specific details on the firmware issues at this point in time.
To prove the successful extraction of private keys, we provide a signature of the paper title created with the extracted keys at~\cite{RobertBuhrenGithub}.
The signature can be verified using certificates provided by AMD.
\end{quote}

The rest of the paper is structured as follows:
In Section \ref{sec:background} we give an overview of the SEV technology, its remote attestation capabilities and the PSP.
Section \ref{sec:firmware_analysis} presents the results of our firmware analysis necessary to understand the attacks.
The attacks are motivated in Section \ref{sec:attack_motivation} and described in Section \ref{sec:attacks}.
Section \ref{sec:discussion} discusses the implications on the security of SEV and proposes a new design for the remote attestation functionality.
In Section \ref{sec:related_work} we give an overview of related work and finally conclude in Section \ref{sec:conclusion}.

%% file: src/2_background.tex
\section{Background}
\label{sec:background}
In this section, we give an overview of basic x86 virtualization concepts and more specifically on the \textit{Secure Encrypted Virtualization} technology by AMD including the Platform Security Processor (PSP).
We focus on the remote attestation feature as this is our main subject of analysis for the rest of the paper.

\subsection{x86 Virtualization Concepts}
\label{subsec:x86_virt}

Hardware extensions to facilitate the use of virtual machines were introduced in 2005 by both Intel (VT-x) \cite{uhlig2005} and AMD (SVM) \cite{svm2005}.
These extensions distinguish between the higher-privileged \emph{host mode} and the lower-privileged \emph{guest mode}.
Both modes are comprised of different privilege \emph{rings}, which allow separating each mode in privileged and unprivileged execution compartments.
The host mode controls the resources, such as memory and CPU time, of the guest mode.

A software component called \emph{hypervisor} executes in the host mode, whereas the guest mode is occupied by the \emph{guest} virtual machines (VMs).
In a cloud scenario, the hypervisor is supplied by the cloud provider and therefore under full control of the cloud provider.

Running in the higher privileged host mode, the hypervisor has full access to the guest VM's memory content.

\subsection{Secure Encrypted Virtualization}
\label{subsec:sev}
The \textit{Secure Encrypted Virtualization} technology (SEV) was first presented by AMD in 2016~\cite{AMD2016}.
It aims to protect a machine in the presence of an untrusted cloud provider.
The primary responsibilities of SEV are runtime protection and secure initialization of virtual machines.

Throughout the rest of the document, the term \textit{platform owner} refers to any entity that owns the SEV platform, i.e., the AMD system that hosts the virtual machines.
The term \textit{guest owner} refers to any entity that intends to utilize the platform provided by the platform owner.
In a typical use case, the platform owner is the cloud computing provider, and the guest owner is a customer of the provider.

The term \emph{SEV firmware} refers to the code running on the SEV platform implementing the SEV API ~\cite{SEVFW}.
It is provided by AMD and hosted by the Platform Security Processor (PSP), in contrast to the main x86 cores (referred to as \emph{main processor} in this document), that are under full control of the platform owner.
\\

\noindent\textbf{\textit{Runtime Protection}}\quad
The core of SEV's runtime protection is a memory encryption engine embedded in the memory controller that encrypts the main memory using AES-128~\cite{AMD2016}\footnote{To protect the guest register state, AMD proposed an extension to SEV: SEV-\textit{Encrypted State} (SEV-ES)~\cite{SEV-ES:2017}.}. 
The encryption keys are generated by a firmware running on a dedicated processor called the \textit{Platform Security Processor} (PSP).
It provides an API~\cite{AMD2018} that the hypervisor, running on the main processor, must use in order to manage the encryption keys for SEV-enabled guests.
An SEV-enabled guest controls which memory pages are passed through the encryption engine by its guest pagetable.\\

\noindent\textbf{\textit{Remote Attestation}}\quad
Even with runtime protection in place, SEV would be useless in the face of an untrusted cloud provider, if the platform owner could modify the VM during deployment or fake the presence of SEV altogether.
Through remote attestation, the SEV firmware implicitly provides proof of authenticity of the SEV platform to the guest owner and explicitly attests a VM's integrity to her.
This is explained in Section \ref{subsec:secure_channel} and \ref{subsec:sev_guest_deployment}.\\

\noindent\textbf{\textit{Platform Security Processor}}\quad
Introduced in 2013~\cite{PSP2013}, the Platform Security Processor (PSP) is a dedicated ARMv7 based co-processor built into the die of AMD CPUs providing security functionality. 
It uses its own memory and non-volatile storage and can access the main processor's system memory. 
The firmware running on the PSP is provided by AMD and integrity protected.

It also hosts a dedicated SEV firmware~\cite{SEVFW} that implements the SEV API specified in~\cite{AMD2018}.

\subsection{SEV: Cryptographic Keys}
\label{subsec:sev_keys}

SEV offers cryptographic proof that a) the remote platform is an authentic AMD platform which supports SEV and b) that a guest was deployed with SEV protection in place. 
To that end, the SEV firmware manages several cryptographic keys that are explained in this section.\\

\begin{figure}[!htb]
  \centering
  \includegraphics[width=\linewidth]{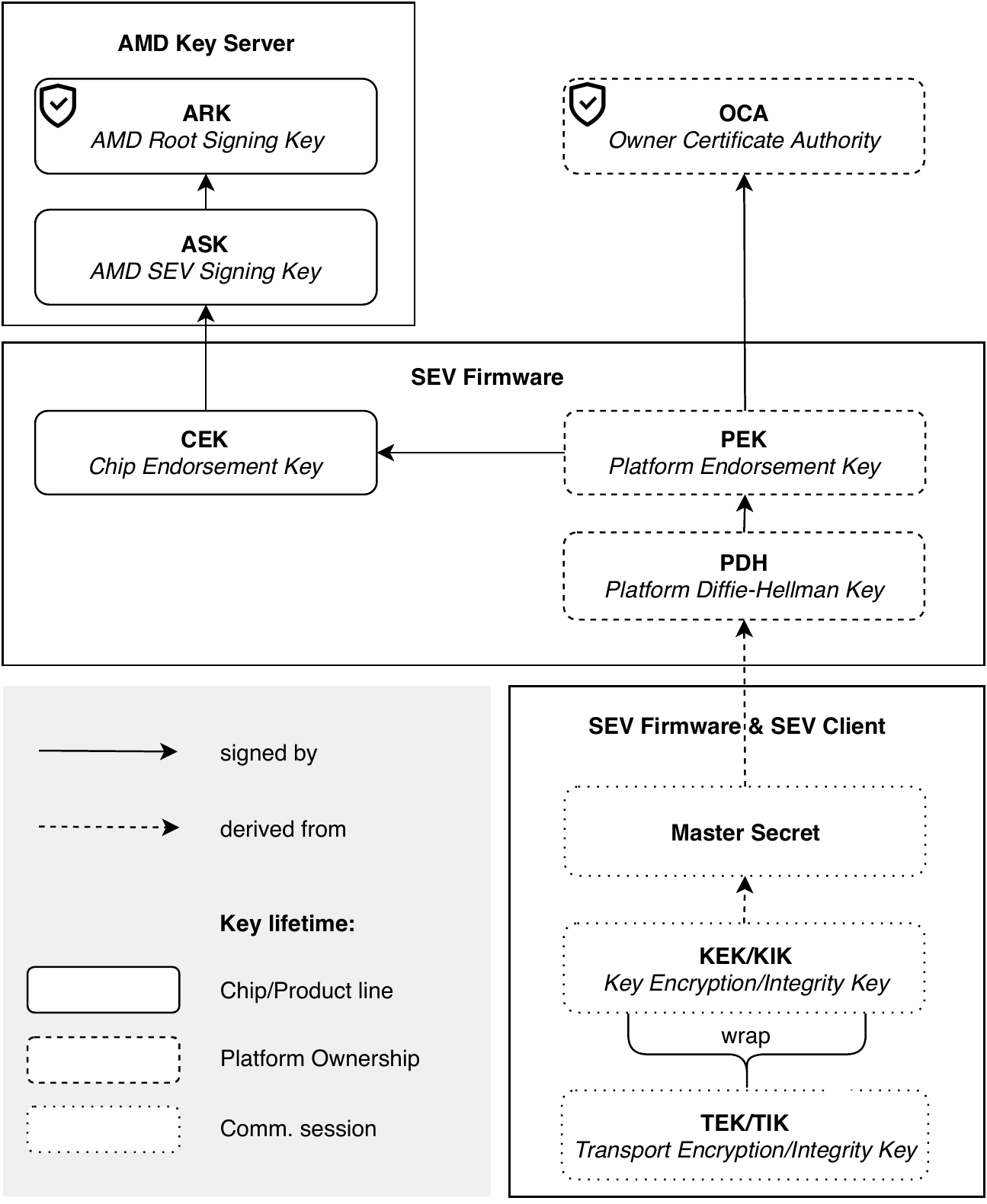}
  \caption{Cryptographic keys in SEV. A shield denotes the key as the root of trust for the corresponding certificate chain. Boxes show the scope of the respective keys.}
  \label{fig:SEV_keys}
\end{figure}

\noindent\textbf{\textit{Firmware Identity}}\quad
Upon initialization, the SEV firmware running on the PSP generates an ECDSA key, the \textit{Platform Endorsement Key} (PEK), using a secure entropy source, see Figure \ref{fig:SEV_keys}.
The SEV firmware uses the PEK to sign the \textit{Platform Diffie-Hellman} key (PDH), which is used to negotiate a shared secret with a remote party, e.g., to establish the secure channel between the guest owner and the SEV platform\cite[Chapter 1.2.2]{AMD2018}.\\

\noindent\textbf{\textit{Platform Ownership}}\quad
Ownership information is provided by signing the PEK with a certificate authority (CA) of the cloud provider, the \textit{owner certificate authority} (OCA), see Figure~\ref{fig:SEV_keys}.
The SEV firmware allows generating a \textit{certificate signing request} (CSR) of the PEK.
This allows the cloud provider to sign the PEK. 
The signed PEK certificate is then re-imported into the SEV firmware~\cite[Chapter 1.2.4]{AMD2018}.\\

\noindent\textbf{\textit{Platform Authenticity}}\quad
To provide the guest owner with an authenticity guarantee of the platform, the PEK is also signed by the \textit{Chip Endorsement Key} (CEK), see Figure~\ref{fig:SEV_keys}.
The CEK is an ECDSA key which is derived from CPU-specific secrets stored in one-time-programmable fuses (OTP fuse) in the CPU~\cite[Chapter 1.2.3]{AMD2018}.
To prove the authenticity of the CEK, it is signed by the \textit{AMD SEV Signing Key} (ASK) which is in turn signed by the \textit{AMD Root Key} (ARK).
As the CEK is unique for each platform, the SEV API specifies a command to retrieve a unique identifier tied to a platform.
While the CEK private key must remain confidential, the signed certificates of the CEK, ASK and the ARK can be obtained from AMD~\cite{AMDKDS} using the platform ID provided by the SEV firmware.
The CEK, therefore, plays a central role in the trust model of SEV.
\\

\noindent\textbf{\textit{Confidential Communication}}\quad
Since SEV assumes an untrusted hypervisor, confidential communication must be ensured for two cases:
First, during the initial deployment of a VM, between the guest owner and the SEV firmware of the target platform.
Second, during migration, between the SEV firmwares of the source and target platform.

To this end, a client and the SEV firmware use the Diffie-Hellman protocol to establish a shared secret, the \textit{master secret}.
A client in this context is either a guest owner or another SEV firmware in case of migration.
Using a key derivation function (KDF) and the established master secret, both the client and the SEV firmware derive the \textit{Key Encryption Key} (KEK) and the \textit{Key Integrity Key} (KIK).
These keys are used to protect the transport keys, the so-called \textit{Transport Encryption Key} (TEK) and \textit{Transport Integrity Key} (TIK).
The transport keys are encrypted using the KEK, and a MAC is generated using the KIK.
This process is referred to as key \emph{wrapping}~\cite[Chapter 2.1]{AMD2018}.
Note that the transport keys are chosen by the client, whereas KIK and KEK are derived from the master secret.

The transport keys are then used to ensure the integrity and confidentiality of data exchanged between the SEV firmware and outside entities.
The following section explains the protocol used to establish the secure channel.

\subsection{SEV: Establish Secure Channel}
\label{subsec:secure_channel}

\begin{figure}[!htb]
  \centering
  \includegraphics[width=\linewidth]{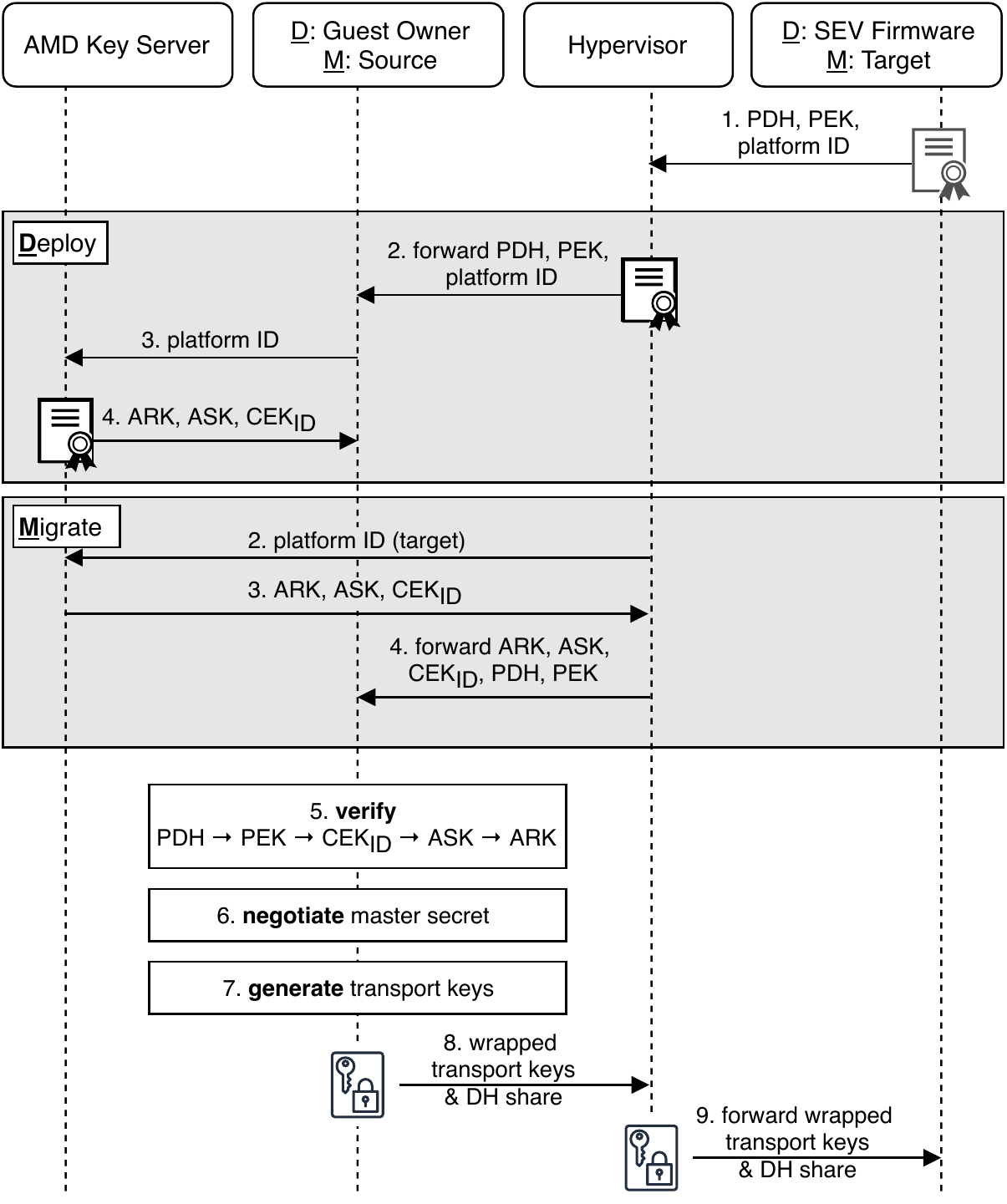}
  \caption{SEV Secure Channel. Protocol initiated by the guest owner, in case of deployment (\underline{D}) and the hypervisor, in case of migration (\underline{M}).}
  \label{fig:secure_channel}
\end{figure}

In order to establish the secure channel, both client and SEV firmware follow the steps depicted in Figure \ref{fig:secure_channel}.
As the SEV API is only accessible via the hypervisor, the secure channel must guarantee authenticity, integrity, and confidentiality of the communication.
In case of deployment, the client is the guest owner, see variant D.
In case of migration, the client is the source platform and the SEV firmware is the target platform of the migration, variant M.

In the first step, the hypervisor retrieves the PDH and the PEK certificate together with a unique platform ID of the target platform.
Using this ID, either the hypervisor, in case of migration, or the guest owner, during deployment, consult the AMD key server to obtain the $CEK_{ID}$ certificate along with the ARK and ASK.
The client is now able to authenticate the target platform by verifying the certificate chain\footnote{To allow the client to verify the ownership of the platform, the PEK can also be signed with the OCA.} as shown in Step 5.
In that case, the client would also verify that the PEK is signed by the OCA in Step 5).
In Step 6 the authenticated PDH is then used to negotiate the master secret using a Diffie-Hellman key exchange~\cite[Chapter 2.2.2]{AMD2018}.
The master secret is only known by the client and the target, but not by the hypervisor.
Using the master secret the key encryption keys are derived, see Section~\ref{subsec:sev_keys}.
Finally, the client generates the transport encryption keys and wraps them using KIK and KEK, Step 7.
The wrapped keys and the Diffie-Hellman share are then transferred to the target.

Both client and the target SEV firmware now hold the transport keys that allow authenticated, encrypted, and integrity protected communication.

\subsection{SEV: Guest Deployment}
\label{subsec:sev_guest_deployment}
While Section \ref{subsec:secure_channel} explained the establishment of the secure channel, this section gives an overview of the steps required to deploy a guest VM in an SEV-enabled cloud system.

Prior to the guest deployment, the platform owner has to initialize the SEV platform. 
During initialization, the SEV firmware derives the platform-specific keys described in Section~\ref{subsec:sev_keys}.
Furthermore, the firmware establishes a chain of trust by signing the PDH with the PEK and the PEK with the CEK:
\begin{equation}\label{eq:1}
  PDH\rightarrow PEK \rightarrow CEK \rightarrow ASK \rightarrow ARK
\end{equation}
This is also depicted in Figure~\ref{fig:SEV_keys}.
Optionally, a second certificate chain is established:
\begin{equation}\label{eq:2}
PDH \rightarrow PEK \rightarrow OCA
\end{equation}
These chains allow a client to authenticate a given PDH\footnote{In case the platform was already initialized, the encrypted and integrity protected state is read from non-volatile storage.}.
After these steps, the SEV firmware transitions into the initialized state.

Before any guest VM can be deployed on the platform, the guest owner authenticates the remote SEV platform.
Using the steps shown in Figure~\ref{fig:secure_channel}, she can establish a secure channel with the remote SEV platform.
The verification of the certificate chain, see Step 5, ensures that the remote system is an authentic AMD system that supports SEV.\\

\begin{figure}[!htb]
  \centering
  \includegraphics[width=\linewidth]{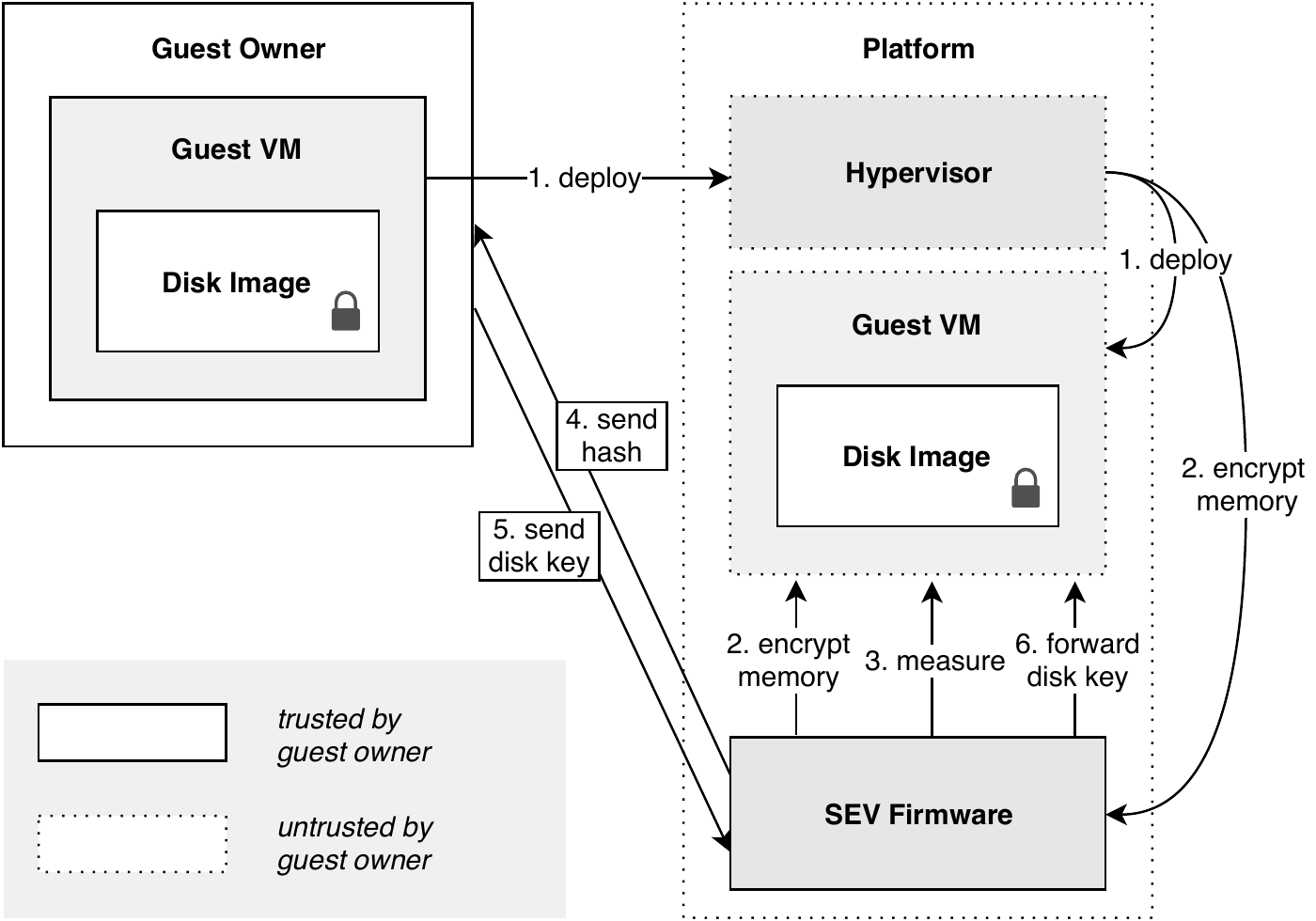}
  \caption{Initial deployment of a guest virtual machine in an SEV scenario.}
  \label{fig:SEV_deployment}
\end{figure}
\noindent\textbf{\textit{Launching the Guest}}\quad \label{subsec:prepare_guest} The guest owner now prepares the guest VM to be executed by the cloud provider.
The initial guest VM is sent to the hypervisor unencrypted and therefore must not contain any confidential data.
To ensure the confidentiality of the guest owner's data, the initial guest image will usually contain an additional encrypted disk image.
In this case, the encryption key is then later provided using the established secure channel.
Besides the VM image itself, the guest owner must also provide a policy that defines restrictions on the actions the cloud provider can perform on the guest VM.
These include, e.g., the cloud provider's ability to migrate the guest VM to another platform or the minimum SEV API version that the target SEV firmware must implement.
As the memory encryption of SEV prevents traditional VM migration, the SEV firmware provides an interface to migrate VMs to a different host, as outlined in the following section.

The guest owner deploys the VM, including the encrypted disk image, to the cloud provider, see Step 1 of Figure~\ref{fig:SEV_deployment}.
The hypervisor launches the guest and calls the SEV firmware to encrypt the memory, Step 2.
Next, the SEV firmware calculates a hash of the initial plaintext VM memory. 
The hash, together with the SEV API version and the guest policy, is protected by the secure channel and transferred to the guest owner, see Step 3 and 4.
The hash gives the guest owner confidence that the VM was deployed unmodified by the hypervisor.
Lastly, in Steps 5 and 6, the guest owner uses the secure channel to provide the disk encryption key to the VM.
This allows the VM to decrypt the disk and process the confidential data.
The guest VM is now fully operational and protected by SEV.

\subsection{SEV: Migration and Snapshots}
\label{subsec:migration_and_snapshots}
Virtual machine migration and snapshotting are common tasks in a cloud computing environment.
Both migration and snapshotting require the export of virtual machine memory.
In case of migration, the memory is then exported to a different platform, while for snapshotting, the memory is saved on the same platform for a later re-import.

As SEV encrypts virtual machine memory using ephemeral keys which never leave the SEV firmware, SEV provides a special mechanism to export memory.
Additionally, the guest owner can impose restrictions on the memory export.
She can prohibit the export using the guest policy, and she can define the minimum SEV API version of the target system using the API fields of the guest policy~\cite[Chapter 3]{AMD2018}.

Using the SEV API, the hypervisor initiates the export of the VM memory on the source platform.
The exported memory is encrypted, and integrity protected using transport keys that are generated by the source SEV firmware.
To that end a secure channel between the source and the target SEV firmware is established, see Figure~\ref{fig:secure_channel} variant M.
The target platform can decrypt the exported memory and re-encrypt it using a freshly generated ephemeral memory encryption key.
This allows the export of encrypted memory in the face of an untrusted hypervisor as only the source and the target SEV firmware share the transport keys that are used to encrypt the exported memory.

%% file: src/3_firmware_analysis.tex
\section{Firmware Analysis}
\label{sec:firmware_analysis}

The Platform Security Processor (PSP) hosts a firmware provided by AMD.
Amongst other things, this firmware implements all SEV related operations carried out by the PSP.
Given the trust model of SEV where the guest owner requests services from an untrusted hypervisor, it is paramount that this firmware is not under control of the platform owner, but instead provisioned by a trusted entity, AMD in this case.
This section presents the results of our firmware analysis on which our attacks in Section \ref{sec:attacks} are based.

\subsection{PSP Firmware Structure}
\label{subsec:psp_firmware_structure}
By analyzing UEFI firmware updates of AMD Epyc systems, we were able to locate the PSP firmware.
It is comprised of several components which are stored in an undocumented area of the UEFI firmware image.
Although the layout of the UEFI firmware residing on the SPI flash is standardized~\cite{UEFISPEC}, the PSP and SEV firmware are not part of the standardized layout.
Building on published information from the Coreboot project~\cite{COREBOOT2014} and the SEV API specification \cite{AMD2018} we were able to understand the proprietary filesystem and identify and extract all firmware components.
The individual firmware components are prepended with a header containing metadata.
This metadata contains a version field and also determines the certificate used to verify the component's integrity.

Most relevant for the attacks presented in this work are three components: The ARK public key (see Figure \ref{fig:SEV_keys}), a component we call \textit{PSP Operating System} (PSP OS) and a component that implements the SEV API, the \textit{SEV firmware}.
To facilitate further research, we developed a PSP firmware analysis tool which is published under~\cite{psptoolGithub}.

PSPTool allows to parse the proprietary filesystem used to store the PSP OS and SEV firmware.
It lists all firmware components alongside various attributes.
Furthermore, it is able to correlate SPI read accesses recorded with a logic analyzer with a given binary.
This allows to inspect the order in which the PSP firmware components are loaded from flash.
A full list of features can be found at~\cite{psptoolGithub}.\\

\noindent\textbf{\textit{ARK Public Key}}\quad
This ARK is a 2048 bit RSA public key stored in a format as described in \cite[Appendix B.1]{AMD2018}.
We could verify that both the PSP OS and the SEV firmware are signed with the ARK private key.
The prepended header of each component is part of the signed data.
The ARK public key is also contained in the ARK certificate that can be obtained from the AMD key server~\cite{AMDKDS}, see Section \ref{subsec:sev_keys}.\\

\noindent\textbf{\textit{PSP Operating System}}\quad
The PSP OS is the only component that contains privileged ARM code. 
It executes in the privileged mode of the PSP, the \texttt{SVC} mode, with paging enabled.
Amongst its responsibilities are system initialization and loading of other firmware components.
These components execute in the unprivileged \texttt{USR} mode and are thus separated from the PSP OS.
There is always only a single unprivileged component present in memory. 
To switch to a different component, the currently loaded component is replaced.
The PSP OS provides a syscall interface for unprivileged components that provides, e.g., access to the \textit{cryptographic co-processor} (CCP).
The CCP is a dedicated hardware component that allows offloading various cryptographic operations. 
It is usable from both the main processor and the PSP.\\

\noindent\textbf{\textit{SEV Firmware}}\quad
The SEV firmware implements the SEV API specification~\cite{AMD2018}. 
It is loaded by the PSP OS and is executed in the unprivileged \texttt{USR} mode.
Its responsibilities include the key generation steps shown in Section \ref{subsec:sev_keys} as well as the policy enforcement described in Step 4 of Section~\ref{subsec:sev_guest_deployment}.
It maintains a state of all SEV-enabled guest VMs as well as the platform state which includes the generated certificates and private keys.
Upon initialization, the SEV firmware will either load a previously saved state from non-volatile storage utilizing the syscall interface provided by the PSP OS or generate a new state including new certificates.
The saved state is encrypted, and integrity protected~\cite[Chapter 5.1.5]{AMD2018}.
The guest state includes the guest policy as well as the guest VM's memory encryption keys.
While the PEK and PDH are generated using a \textit{``secure entropy source''} ~\cite[2.1.1f)]{AMD2018}, the CEK is derived from \textit{``chip-unique OTP fuses''} and has a lifetime of the corresponding CPU~\cite[2.1.3]{AMD2018}.
To retrieve the value from the OTP fuse, the SEV firmware issues a syscall.
The PSP OS will then retrieve 32 bytes from the CCP and relay it to the SEV firmware. 
These 32 bytes are used as an input to a key derivation function.

\subsection{PSP Boot Security}
\label{subsec:boot_security}
To better understand the PSP boot process, we used a logic analyzer to record accesses to the SPI flash memory hosting the UEFI image.
The PSP firmware components are stored alongside the UEFI firmware.
Specifically, the PSP firmware resides in UEFI padding volumes (Pad File), see~\cite[Chapter 2.1.4]{UEFISPEC}.
We observed that the first components that are loaded from flash are the ARK public key and the PSP OS.
We also observed a delay after the PSP OS is loaded and before any other attempt to access the flash can be observed.
Our experiments have shown that a modified ARK will result in no further flash reads after the ARK is read.
In case the PSP OS was altered, a PSP OS from a different flash location is loaded.
If this recovery PSP OS is also altered, the system resets.

\begin{figure}[h]
  \centering
  \includegraphics[width=\linewidth]{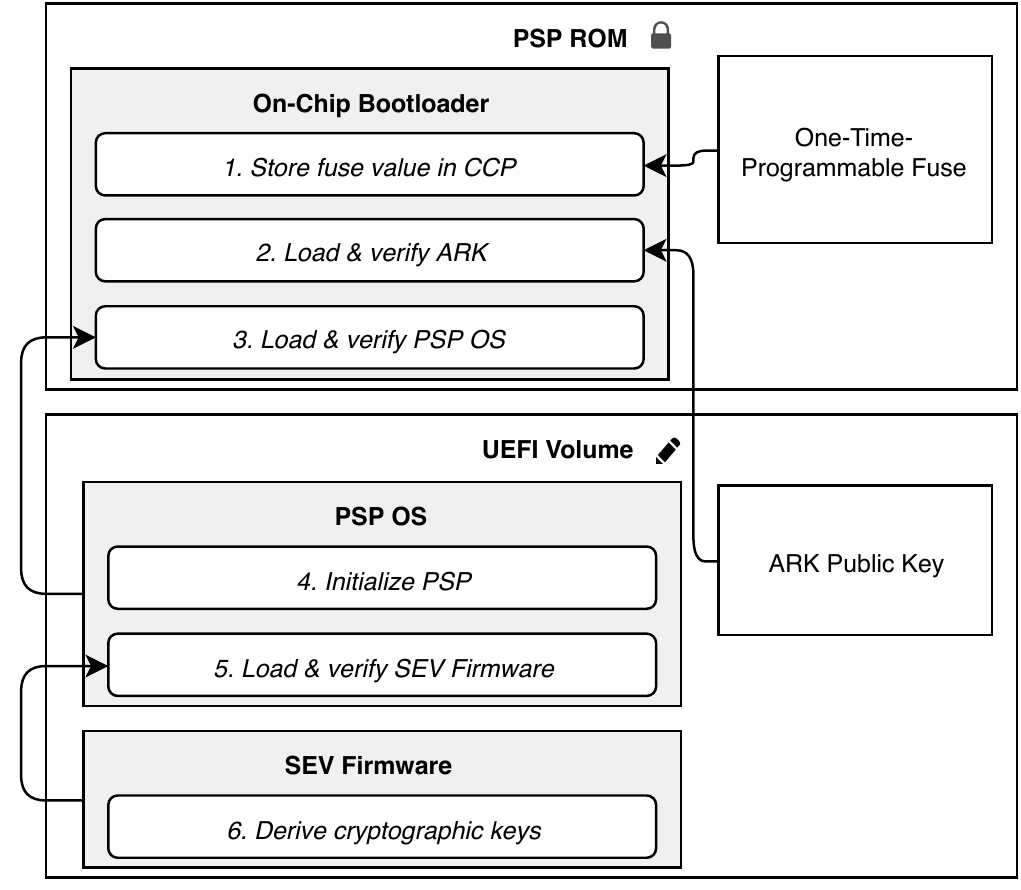}
  \caption{Boot procedure on an SEV-enabled system. A lock denotes a non-modifiable component whereas a pen denotes modifiable components. Arrows denote dependencies of initialization steps. }
  \label{fig:PSPboot}
\end{figure}

Based on these observations and our static analysis from Section \ref{sec:firmware_analysis}, we inferred a boot order, as shown in Figure \ref{fig:PSPboot}.
The Figure focuses on the security-relevant parts of the SEV technology; other steps of the boot process are omitted.

In addition to the PSP OS and the SEV firmware, a third component is responsible for bootstrapping the system.
This component is responsible for loading the ARK certificate and the PSP OS from flash, steps 2 \& 3.
Since it is not part of the PSP firmware loaded from flash and we conclude that it is stored in a read-only memory (ROM) of the PSP, we call it \emph{on-chip bootloader}.

Our static analysis revealed that the OTP value that is used to derive the CEK, see Section \ref{subsec:sev_keys}, is provided by the PSP OS via a syscall. 
The implementation of that syscall simply forwards a value from the CCP to the calling unprivileged component. 
This value is stored in a \emph{storage block} of the CCP.
These storage blocks are used by the CCP to maintain a context for cryptographic routines~\cite{CCPDEV}.
From that, we conclude that the on-chip bootloader is also responsible for storing the OTP value in the CCP storage block, see Step 1 of Figure \ref{fig:PSPboot}.

After the integrity of the PSP OS is verified, the PSP OS is executed.
It initializes the PSP, Step 4, which includes basic operating systems tasks such as initializing the pagetables and peripherals, e.g., the interrupt controller.
In Step 5, it loads the SEV firmware and verifies its integrity using the ARK public key. 
The ARK public key is not loaded from flash again, but instead, the PSP OS assumes the presence of that key at a fixed memory location.
We infer that the PSP on-chip bootloader placed the ARK public key there after its integrity was verified in Step 2.
Lastly, upon initialization of the SEV firmware, the firmware either loads the platform state from non-volatile storage, see Section~\ref{subsec:sev}, or generates the cryptographic keys, Step 6, as described in Section~\ref{subsec:sev_guest_deployment}.
\\

\noindent\textbf{\textit{Security Issues}}\quad
Our experiments revealed a security issue in the signature verification mechanism of the PSP OS which allowed us to load custom components through modified UEFI images.
Either the attacker has physical access to the SPI flash chip and makes use of an SPI programmer, or the UEFI update mechanism of the motherboard vendor allows her to flash a custom image. 
Depending on the vendor, there is no signature check on the UEFI image in place~\cite{SUM} or it can be disabled~\cite{AMIAFU}.
These tools can be used remotely, so no physical access to the target device is required.
Similar issues were previously published by security researchers~\cite{CTSLABS2019}.
AMD confirmed the existence of these issues~\cite{AMDBLOG} and newer versions of the PSP OS are already patched.
However, our experiments showed that the PSP does not employ any rollback prevention mechanism, i.e., even though a system already makes use of a patched PSP OS, it is always possible to revert to an earlier, vulnerable version.
Thus, as the issues exist in signed versions of the PSP OS, mitigations require changes to the component verifying the signature, the on-chip bootloader, in order to prevent a vulnerable PSP OS version from being loaded.

AMD confirmed to us in personal correspondence that the on-chip bootloader is not updatable through a firmware update. 
Instead, using the UEFI update mechanism, the PSP is able to update a fuse configuration that allows changing the key that is used to authenticate the PSP firmware components.
This could be leveraged to revoke the currently used key and thus prevent the vulnerable PSP OS version to be loaded.
In this case, updated versions of the PSP OS would have to be signed with an alternative key.

But even though multiple issues of the PSP OS have been reported to AMD in mid 2018~\cite{CTSLABS2019}, we have not found any evidence of a key revocation.
To that end, we analyzed ten different binaries from five different motherboard vendors.
At the time of writing, the latest UEFI updates contain PSP OS versions that are signed with the same key as the vulnerable PSP OS version.
From that, we conclude that AMD has not revoked any ARK public keys in order to suspend vulnerable PSP firmware versions.
To the best of our knowledge, the same ARK is used across all CPUs of the AMD Epyc Naples series.
This indicates that every CPU of that series is affected as any PSP OS version is loaded as long as it is signed with the ARK.
Based on these results we conclude that for the AMD Epyc Naples series CPUs, an attacker is able to:
\begin{enumerate}
  \item \textbf{Execute custom code on the PSP} using a vulnerability in the signature verification mechanism of the PSP OS.
  \item \textbf{Roll back} from any PSP OS version to a \textbf{vulnerable PSP OS} version.
\end{enumerate}

\subsection{CEK Extraction} 
\label{sub:cek_extraction}
Leveraging the security issues discussed in the previous section, we built and deployed a patched SEV firmware.
This firmware allows us to read and write arbitrary PSP memory.
Using this patched SEV firmware, we extracted the SEV state, see Section~\ref{subsec:psp_firmware_structure}, including the CEK private key of three different AMD Epyc CPUs.

We obtained the corresponding signed CEK certificates from the AMD key server, see Section~\ref{subsec:sev_keys}, and verified the extracted private keys by creating signatures that can be validated using the signed CEK certificate.
A proof-of-concept signature created with an extracted CEK can be found at~\cite{RobertBuhrenGithub}.

Although the exact details of the CEK extraction are omitted in this paper, we will provide security researchers with additional details to reproduce our results upon request.
We plan to release the exact details once a fixed hardware platform is available for customers.

As both the authenticity of the SEV platform as well as the confidentiality of the data protected by SEV rely on the security of the CEK, the CEK extraction lays the groundwork for the first two attacks in Section \ref{sec:attacks}.
The motivation for our attacks is discussed in the following section.

%% file: src/4_attack_motivation.tex
\section{Attack Motivation}
\label{sec:attack_motivation}

The SEV technology offers data protection of virtual machines in the face of an untrusted platform owner.
This incorporates a malicious cloud provider or a malicious system administrator.
In this section, we present two different motivations for an attacker to circumvent the security properties of SEV as presented in Section \ref{sec:background} and Section \ref{sec:firmware_analysis}.

\subsection{Extract Confidential Data} \label{subsec:extract_confidential_data}

The first motivation regards data theft and originates from an individual targeting an SEV guest owner.

The additional security measures provided by SEV enable companies to process confidential data in the cloud, that would otherwise not be processed in the cloud because of its confidentiality.
The goal of the first type of attacker is to get access to this data despite the presence of SEV.

An individual with malicious intent and sufficient permissions could use data from a commercial guest owner to pursue traditional fraud, e.g., with stolen credit card data.

\subsection{Save Resource Overhead} \label{subsec:minimize_mem}

The second motivation is of economic nature and originates on the organizational level of a cloud provider.

In an \textit{Infrastructure-as-a-Service} (IaaS) scenario, the cloud provider charges guest owners based on the amount of resources they allocate, including CPU, memory, and disk utilization.

In order to increase the overall utilization of memory, many hypervisors employ \textit{Kernel same-page merging} (KSM)~\cite{arcangeli2009increasing} to increase the memory utilization.

KSM requires the hypervisor to read the guest memory of virtual machines in plaintext to identify duplicate pages. 
Using SEV prohibits KSM, since memory pages of different guest VMs are encrypted using different keys.
Therefore, the memory requirement of an SEV-enabled system is increased, which results in higher costs for the cloud provider.
In an IaaS scenario, it is likely that the cloud provider will pass those additional costs to the customers commissioning the security features of SEV.

While a benevolent cloud provider might disable KSM to prevent attacks such as~\cite{yarom2014flush+}, and \cite{razavi2016flip}, this is not necessarily the case for a provider with malicious intent.

In order to increase revenue, a malicious cloud provider could fake the presence of SEV, while still charging additionally for SEV protection.
The guest VM would instead be hosted traditionally on a non SEV-enabled system, leveraging, e.g., KSM to reduce the memory consumption and therefore the costs for the cloud provider.

%% file: src/5_attack.tex
\section{Attacks}
\label{sec:attacks}

In this section, we propose three different attacks targeting AMD SEV.
These are structured according to the two main attack motivations, as described in Section \ref{sec:attack_motivation}.

\subsection{Fake SEV}
\label{subsec:fake_sev_attack}

As presented in Section \ref{subsec:minimize_mem}, SEV prevents the use of virtualization features like KSM, thus increasing the overall memory requirements of a cloud setup. 
This additional cost may motivate a malicious cloud provider to fake the presence of SEV.
Furthermore, by faking the presence of SEV, the cloud provider gains access to data which is not accessible when protected by SEV.
Using an extracted CEK private key as shown in Section \ref{sub:cek_extraction}, the cloud provider may pose as an authentic AMD SEV platform even though SEV is not enabled or even present at all.
\\

\noindent\textbf{\textit{Attack model}}\quad
The attacker is a cloud provider running arbitrary hosting hardware, who has had access to any SEV-enabled system for one-time extraction of the CEK private key and the corresponding platform ID.
It is not required that this specific system hosts the victim's VM.
The victim is a cloud customer expecting an SEV-enabled VM from the attacker.
This attack does not impose restrictions on the guest system in use.
\\

\noindent\textbf{\textit{Method}}\quad
From the point of view of a guest owner, the correct deployment of VMs can be verified through remote attestation by the trusted SEV firmware.
As described in Step 5 of Figure \ref{fig:secure_channel}, the guest owner authenticates the remote platform using a certificate chain (as seen in Figure \ref{fig:SEV_keys}) that roots in the AMD Root Signing Key (ARK).
Once the remote platform is authenticated, the guest owner defines the transport keys, wraps them, and sends them to the cloud provider, see Steps 7 and 8 of Figure~\ref{fig:secure_channel}.

In order to simulate the presence of SEV, the attacker first generates an arbitrary PEK and, if required, signs it using the OCA.
Next, she uses the extracted CEK private key (see Section \ref{sub:cek_extraction}) to sign the PEK. 
She then generates a PDH which is in turn signed by the PEK.
Now, the attacker has control over the highlighted parts of the certificate chains:
\begin{equation}
  \tcboxmath[size=small]{PDH} \rightarrow \tcboxmath[size=small]{PEK} \rightarrow \tcboxmath[size=small]{CEK} \rightarrow ASK \rightarrow ARK
\end{equation}
\begin{equation}
  \tcboxmath[size=small]{PDH} \rightarrow \tcboxmath[size=small]{PEK} \rightarrow \tcboxmath[size=small]{OCA}
\end{equation}
\noindent To mount the attack, the attacker provides the platform ID corresponding to the CEK, the PEK, and PDH to the guest owner.
The guest owner will obtain the ASK, ARK, and CEK certificates from the AMD key server, see Steps 3 and 4, variant D of Figure~\ref{fig:secure_channel}.
After verification of the certificate chains, the guest owner deploys the guest VM, including the encrypted transport keys.
As the attacker possesses the private PDH key, she can decrypt the transport keys provided by the guest owner.
Now the attacker calculates a hash of the guest VM's memory.
As shown in Figure~\ref{fig:SEV_deployment} Step 4, the hash is provided to the guest owner.
Due to the fact that the attacker controls the transport keys, she can provide the hash of the guest VM's memory and protect it using the transport keys.

The last step in the guest deployment phase is to provide a secret, e.g., a disk encryption key, to the guest VM, see Step 5 of Figure~\ref{fig:SEV_deployment}.
The guest owner protects the disk encryption key using the transport keys.
As they are known to the attacker, she can first decrypt the disk encryption key and then decrypt the encrypted disk to get access to the guest virtual machine's confidential data.
To fake the presence of SEV, the attacker injects the encrypted disk encryption key into the guest virtual machine, i.e., she copies it into the guest's memory.
The guest is now fully operational; however, although the remote attestation mechanism of SEV was successfully carried out, the guest owner has no means to detect the absence of the SEV feature.

This enables a malicious cloud provider to increase the number of guests on a single host making use of KSM, see Section~\ref{subsec:minimize_mem}.
More so, as the runtime protection of SEV is not enabled, the attacker can access any data used in the guest VM.

\subsection{Migration Attack}
\label{subsec:migration_attack}

As outlined in Section \ref{subsec:extract_confidential_data}, the goal of this attack is to extract run-time data of an SEV-enabled guest from a host system.
\\

\noindent\textbf{\textit{Attack model}}\quad
The attacker is an individual, e.g., a system administrator of an otherwise trusted organization, with access to the management interface of an SEV-enabled host.
The victim is a cloud customer who successfully deployed a virtual machine on the SEV-enabled host of the cloud provider.
We assume that there are no security issues present in the guest VM and the PSP firmware on the host.
The attacker must have had access to any SEV-enabled system for one-time extraction of the CEK private key and to obtain the corresponding CEK certificate.
It is not required for the attacker to have access to the CEK private key belonging to the platform hosting the VM.
Furthermore, he must be able to initiate a virtual machine migration of the victim's VM using the management interface of the host.
Additionally, the guest policy must allow migration of the guest.
To benefit from cloud services such as high availability and dynamic resource allocation, migration is required in order to handle resource contention or failures in the host system.
Thus it is likely that the guest policy allows migration.
\\

\noindent\textbf{\textit{Method}}\quad
Similarly to the previous attack, the attacker first needs to get hold of a valid CEK private key of any authentic SEV-enabled system, see Section \ref{sub:cek_extraction}.
The attacker creates the two certificate chains as described in Section~\ref{subsec:fake_sev_attack}, chain 3, and chain 4.

Using the SEV API commands for VM migration (see Section \ref{subsec:migration_and_snapshots}), the attacker instructs the SEV firmware to initiate the migration of the victim's VM using the prepared PDH, PEK, and ARK.
Using the attacker-controlled PDH, the SEV firmware on the source host will authenticate the target platform of the migration, see Step 5 of Figure~\ref{fig:secure_channel}.
Since the provided certificates were created using a valid (extracted) CEK, the SEV firmware will accept the PDH.
The SEV firmware then generates the transport keys, and wraps them using keys derived from the authenticated PDH, see Steps 7 and 8 of Figure~\ref{fig:secure_channel}.
The memory of the VM is encrypted using the generated transport keys and exported along with the wrapped transport keys.

Instead of forwarding the keys, see Step 9 of Figure~\ref{fig:secure_channel}, the attacker can now unwrap the transport keys and decrypt the virtual machine's memory since he controls the PDH that was used to derive the keys used to encrypt the transport keys.

We emphasize that this attack does not require any security issues to be present in the PSP firmware of the source host.
By owning any CEK private key, the attacker can impersonate a valid target for migration.
As the transport encryption keys in SEV must be shared with the target of a migration, the attacker, posing as a valid migration target, can decrypt the exported memory.

\subsection{Debug Override Attack}
\label{subsec:debug_override_attack}

Similarly to the previous attack and as outlined in Section \ref{subsec:extract_confidential_data}, the goal of this attack is to extract run-time data of an SEV-enabled guest from a host system.
\\

\noindent\textbf{\textit{Attack model}}\quad
The attack model is similar to the Migration Attack in Section~\ref{subsec:migration_attack}, with the additional requirement that the attacker must be able to install UEFI updates on the system that hosts the victim's VM.
UEFI updates can be deployed remotely without any physical access using the server's update mechanism, see Section~\ref{subsec:boot_security}. 
Alternatively, they can be deployed with physical access via directly programming the SPI flash located on the server's motherboard using an SPI programmer.
\\

\noindent\textbf{\textit{Method}}\quad
The SEV API specifies a debug interface to assist debugging of SEV-protected virtual machines.
The debug interface allows a hypervisor to read and write guest memory in plaintext, see \cite[Chapter 7.1]{AMD2018}.
For example, in a QEMU/KVM scenario, QEMU offers the \textit{pmemsave} command to dump guest memory to a file~\cite{qemuDoc}.
In an SEV-enabled guest, that memory is encrypted and thus of no use to debug the guest.
The SEV API debug commands enable the hypervisor to dump plaintext memory instead.
The SEV firmware will only allow the use of this interface if the guest owner explicitly enabled debugging in the guest policy as described in Section \ref{subsec:prepare_guest}.

Using the security issues described in Section~\ref{subsec:boot_security}, an attacker is able to patch the SEV firmware so that it ignores the policy.
After installing the patched firmware, the SEV firmware will decrypt or encrypt guest memory regardless of the guest owner's policy.
This allows the attacker to read and write arbitrary guest memory using the SEV debug interface from the SEV-enabled host.

Leveraging these security flaws, we were able to successfully install such a patched version of the SEV firmware.
As opposed to previous attacks on SEV such as \cite{DBLP:journals/corr/abs-1805-09604} and \cite{Hetzelt:2017}, this attack does not depend on any services running inside the guest VM.

This attack is also possible if the firmware vulnerabilities described in Section \ref{subsec:boot_security} are fixed.
Due to the missing rollback prevention, an attacker can always replace the existing PSP OS with a vulnerable version before installing her patched SEV firmware.

%% file: src/6_discussion.tex
\section{Discussion}
\label{sec:discussion}
In the previous sections, we laid out how security issues in the PSP firmware pave the way for attacks against SEV that permanently break the security properties of the SEV technology on AMD Epyc based systems.
Furthermore, we demonstrated that, although the issues have been addressed through firmware updates, the confidentiality of SEV-protected systems is still at risk.
This is due to the fact that the presence of an updated firmware cannot be verified by a guest owner.
A given CEK is valid throughout the lifetime of the CPU and does not depend on the firmware version.
This section discusses possible mitigations and proposes a key generation design for future SEV implementations.

Without vulnerabilities as the one described in Section \ref{sec:firmware_analysis}, none of our attacks would be possible. 
Various methods such as static and dynamic analysis, formal verification, and extensive code audits naturally come to mind when discussing possible mitigations. 
However, the sheer amount of security issues that are present in related systems, such as Intel ME~\cite{INTEL-SA-00185, INTEL-SA-00191} and the PSP itself~\cite{CTSLABS2019}, prove the difficulty to ensure the absence of security issues.
We hence believe that the security design of SEV should incorporate the possibility of firmware bugs.
Thus the proposed mitigations focus on SEV \textit{design} changes that empower a guest owner to enforce the use of both an authentic as well as an \emph{up-to-date} firmware on the remote platform.

\subsection{Current Design Issues}
\label{subsec:design_issues}
As discussed in Section~\ref{subsec:boot_security}, the PSP allows installing any signed firmware, including rollbacks to previous insecure versions.
This allows attackers to provision any AMD Epyc CPU with a vulnerable PSP OS version and mount the attacks discussed earlier.
While the current on-chip bootloader is not updatable, AMD confirmed that it is possible to revoke an ARK and enforce the use of alternative keys to verify the integrity of the PSP OS.
This mechanism can be used to label every currently available PSP firmware as untrusted, i.e., effectively preventing an attacker to roll back to a vulnerable PSP OS version.
However, it does not allow a guest owner to verify that a PSP OS version signed with the alternative key is actually used.
Furthermore, the CEK does not depend on the PSP OS version, i.e., a CEK extracted before the revocation of the ARK key is still valid afterwards - its lifetime is the lifetime of the CPU~\cite[Chapter 2.1.3]{AMD2018}.
Thus this approach is insufficient to mitigate our proposed attacks.
This is also true for another SEV mechanism: SEV allows the guest owner to enforce the SEV API version implemented by the remote platform, see Section~\ref{subsec:sev_guest_deployment} Step 4.
However, the SEV API check is only enforced by the SEV firmware.
An attacker able to manipulate the SEV firmware can spoof arbitrary SEV API versions.

The validity of the CEK across firmware versions impedes mitigations only based on firmware updates.
To overcome this, we propose design changes to the SEV technology which are laid out in the next section.

\subsection{Proposed SEV Design Changes}
\label{subsec:design_changes}
The goal of our proposed design changes is to enable the guest owner to enforce the use of an authentic and \emph{up-to-date} PSP firmware.
It is specifically not our goal to provide means to ensure the PSP firmware has no bugs in the first place. 
We rather focus on design changes that allow ensuring the PSP firmware in use is still trusted. 
Our proposed changes aim to incur only low complexity overhead to the current SEV technology.
This enables re-use of the current software stack and minimizes the required effort to migrate to our proposed design.\\

\noindent\textbf{\textit{CEK Derivation}}\quad
The current CEK is derived using a key derivation function (KDF) that takes a 32-byte secret value which is unique per CPU and is stored in \emph{one-time-programmable} (OTP) fuses ($S_{OTP}$):
\begin{equation*} \label{equ:original_cek_kdf}
   \large
   CEK = \mathrm{KDF}(S_{OTP})
 \end{equation*}

We propose to change the way the CEK is generated in order to connect it to the PSP OS version and the SEV firmware version. To that end, we introduce a two-stage secret generation procedure.
Instead of deriving the CEK directly from the $S_{OTP}$, two intermediate secrets are derived using different inputs:

\begin{enumerate}
  \item $S_{PSP}$: based on the PSP OS version ($PV$) and $S_{OTP}$.
  \begin{equation*} \label{equ:s_psp}
   \large
   S_{PSP} = \mathrm{KDF}(PV, S_{OTP})
 \end{equation*}

  \item $S_{CEK}$: based on the SEV firmware version ($SV$) and $S_{PSP}$.
  \begin{equation*} \label{equ:s_cek}
   \large
   S_{CEK} = \mathrm{KDF}(SV, S_{PSP})
\end{equation*}
\end{enumerate}

\noindent The final CEK is then derived from $S_{CEK}$:
\begin{equation*} \label{equ:proposed_cek_kdf}
   \large
   CEK = \mathrm{KDF}(S_{CEK})
\end{equation*}

\noindent For the sake of simplicity, static, i.e., non-confidential, inputs to the KDF were omitted.
The resulting CEK will not only depend on the chip-unique $S_{OTP}$, but also on the current PSP OS version as well as the SEV firmware version.
As the intermediate secrets must not be accessible by an attacker, our design separates the derivation of those secrets in different PSP firmware components.

\begin{figure}[h]
  \centering
  \includegraphics[width=\linewidth]{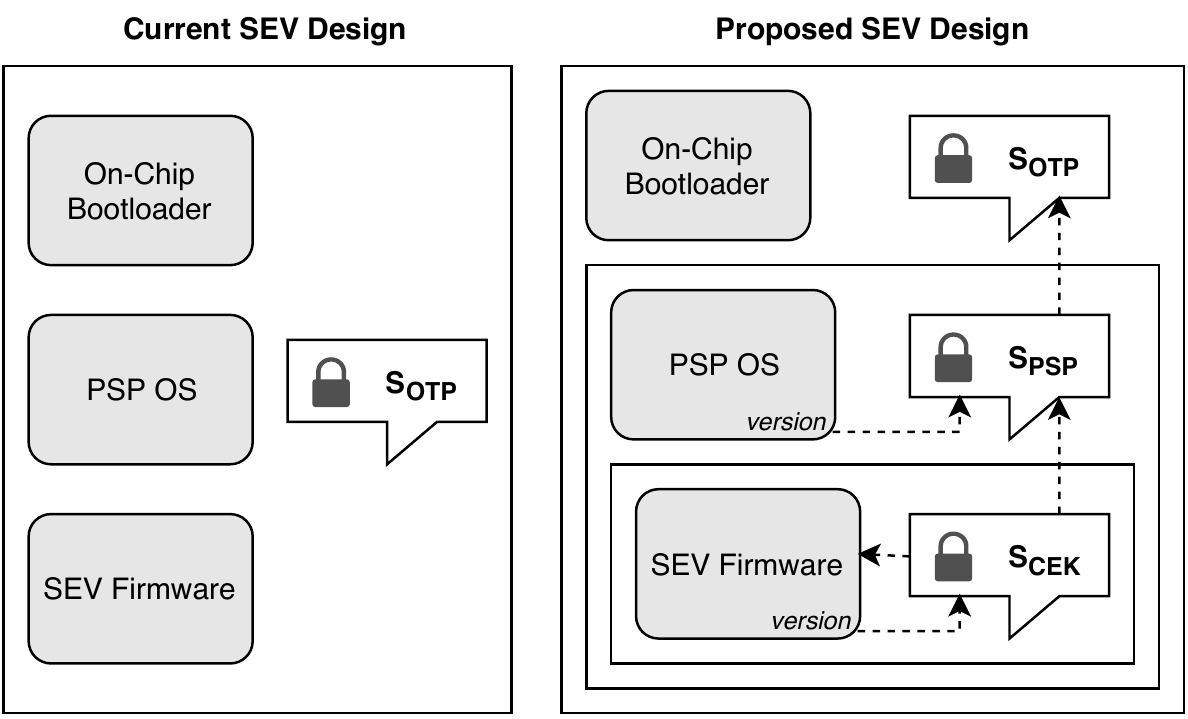}
  \caption{The current SEV design as opposed to our proposed design. Dashed lines between two secrets show a derivation. Boxes show the scope of secrets and firmware components.}
  \label{fig:proposed_changes}
\end{figure}

In the current SEV design, the initial CEK secret, $S_{OTP}$, is accessible to the on-chip bootloader, the PSP OS as well as the SEV firmware (see left part of Figure \ref{fig:proposed_changes}).

Our new design proposes a better isolation, as depicted on the right hand side of Figure \ref{fig:proposed_changes}.
The $S_{OTP}$ is only accessible to the on-chip bootloader and is used in conjunction with the PSP OS version to derive the $S_{PSP}$. 
The $S_{PSP}$ is accessible by the PSP OS and is given as an input to a key derivation function together with the SEV firmware version.
The derived $S_{CEK}$ is used by the SEV firmware to finally derive the CEK.

The required changes in the firmware components are discussed in the following paragraphs.\\

\noindent\textbf{\textit{On-chip Bootloader}}\quad
The current on-chip bootloader provisions the Cryptographic Co-Processor (CCP) with the $S_{OTP}$, see Step 1 of Figure \ref{fig:PSPboot}.
In our proposed design, the on-chip bootloader provisions the CCP with $S_{PSP}$ instead.

$S_{PSP}$ is derived from the $S_{OTP}$ and the PSP OS version using a key derivation function.
The PSP OS version is a field in the signed header of the PSP OS component stored on flash. 
The current on-chip bootloader is already required to read this header as it contains information about the key used to verify its signature.
In addition, the on-chip bootloader needs to implement the key derivation function that derives the $S_{PSP}$.
It is crucial for our enhanced design, that the original $S_{OTP}$ is never visible outside the scope of the on-chip bootloader.
To that end, the hardware component implementing the access to the $S_{OTP}$ must prevent further accesses after the first access.

As the on-chip bootloader is not updatable, it is paramount that its code complexity is rather low.
Any errors in this component that are identified after the CPU is manufactured and shipped cannot be fixed.

While our proposed changes do increase the overall complexity of the on-chip bootloader, we believe they are reasonable and manageable.
Determining the PSP OS version incurs only little overhead as this information is present in the PSP OS header stored on flash.
Including the PSP OS version in the $S_{PSP}$ simply requires parsing one additional header field.

The key derivation function does add additional complexity.
However, the PSP system has access to the CCP, which offers the possibility to offload cryptographic operations. 
We believe that the complexity of the KDF implementation can be reduced by offloading the cryptographic primitives, such as hash functions, to the CCP.
In fact, the CCP is already leveraged to verify the signature of the firmware components.
The proposed KDF in the on-chip bootloader could make use of the CCP in a similar fashion.

Limiting the privilege of accessing the $S_{OTP}$ to only the on-chip bootloader effectively reduces the risk of leaking the $S_{OTP}$ as Figure \ref{fig:proposed_changes} illustrates.\\

\noindent\textbf{\textit{PSP OS}}\quad
In contrast to the original SEV design, the PSP OS must not get access to the original $S_{OTP}$. 
Instead, it only has access to the intermediate secret $S_{PSP}$ that depends on the PSP OS version.
Similarly to the proposed on-chip bootloader changes, the PSP OS uses the SEV firmware version together with the $S_{PSP}$ to derive the $S_{CEK}$.
The SEV firmware version information is present in the header of the SEV firmware which is parsed by the PSP OS.
The resulting $S_{CEK}$ now depends on the $S_{OTP}$, the PSP OS version and the SEV firmware version.
As the PSP OS runs at a higher privilege level (\texttt{SVC} mode) than the SEV firmware (\texttt{USR} mode), the intermediate $S_{PSP}$ is not accessible by the SEV firmware.
Only the $S_{CEK}$ is provided to the SEV firmware through a syscall.\\

\noindent\textbf{\textit{SEV API}}\quad
To accommodate for our proposed design changes, the format of the CEK certificate, see \cite[Appendix C.1]{AMD2018}, must be extended.
In the current SEV API, version 17 at the time of writing, the CEK certificate contains no information about the PSP components.
We propose to extend the CEK certificate format to also include the minimum PSP OS and SEV firmware versions.
As opposed to the original SEV design, there are now multiple valid CEKs for a single CPU.

To enable the guest owner to enforce firmware versions of the remote platform, we further propose to extend the guest policy to include the PSP OS version as well as the SEV firmware version.\\

\noindent\textbf{\textit{SEV Firmware}}\quad
The current SEV firmware derives the CEK from a secret value that is provided by the PSP OS.
This does not change in our proposed design. 
However, the $S_{OTP}$ is not exposed to the SEV firmware. 
Instead, the CEK is derived from the $S_{CEK}$ which is accessible via a syscall.

As there are now multiple CEKs for a single platform, the SEV API must support to enforce the minimum SEV firmware version and PSP OS version that is defined by the guest policy.

During the initial deployment, the guest owner will additionally receive the firmware versions, Step 1 of Figure~\ref{fig:secure_channel}, and will query the AMD key server using the platform ID together with the stated firmware versions, see variant D, Step 3 of Figure~\ref{fig:secure_channel}.
The retrieved CEK is then used in the certificate chain verification, as shown in Step 5 of Figure~\ref{fig:secure_channel}.
The verification will only succeed if the remote platform hosts the stated firmware versions.

In case of migration, the source SEV firmware enforces the minimum firmware versions when authenticating the target platform, Step 5 of Figure~\ref{fig:secure_channel}.
To that end, the source SEV firmware verifies that the versions specified in the provided CEK certificate of the target, see variant M, Step 4 of Figure~\ref{fig:secure_channel}, are equal to or greater than the versions specified in the guest policy.\\

\noindent\textbf{\textit{AMD Key Server}}\quad
In the current SEV design, the AMD Key Server provides means to retrieve a CPU-specific CEK certificate for a given platform ID, as shown in variant D, Step 3 of Figure~\ref{fig:secure_channel}.
In our enhanced design, the AMD Key Server is queried with a platform ID as well as PSP OS and SEV firmware versions.

In a similar fashion to the proposed key derivation introduced above and based on the CPU-specific $S_{OTP}$, the AMD Key Server will calculate the intermediate secret $S_{PSP}$ before delivering the version-dependent $S_{CEK}$ to the client.
It is left to AMD how to communicate the revocation of certain PSP OS and SEV firmware versions for use with the SEV technology.
One possible option is to disable the generation of CEK certificates for known vulnerable PSP OS and SEV firmware versions.
We believe that the on-demand calculation of CEKs increases the required computational effort for the AMD Key Server to a manageable degree.

\subsection{Security Evaluation}
This section discusses the advantages of our proposed design.
For the following paragraphs, we assume that a previously released PSP firmware version contained security issues which are fixed in a later version.
This is the case for the current state of the PSP firmware~\cite{AMDBLOG}.
Furthermore, we assume that AMD publishes information about the outdated, vulnerable firmware versions along with the updated version.
Additionally, we assume that it is not possible to extract the CEK from the current PSP firmware version.\\

\noindent\textbf{\textit{Guest Deployment}}\quad
The proposed SEV design changes allow a guest owner to enforce the use of specific firmware versions on the remote platform.
As the CEK is now tied to the firmware version deployed on the remote platform, a CEK extracted using different firmware versions is no longer valid. 
In the current SEV design, the security of the SEV technology itself is compromised as long as a bug exists in any relevant PSP firmware component.

During the deployment of a guest VM, the guest owner authenticates the remote platform, see Step 5 in Figure \ref{fig:secure_channel}.
In our enhanced design, the guest owner retrieves the CEK certificate for the up-to-date firmware, i.e., the firmware version that includes the bug fixes.
To that end, she uses the platform ID of the target platform along with the required PSP OS and SEV firmware version to retrieve the CEK certificate from the AMD key server, variant D, Step 3 and 4 of Figure~\ref{fig:secure_channel}.

While a malicious cloud provider could provide an extracted CEK from an outdated, vulnerable firmware, the CEK will not match the CEK certificate served by the AMD Key Server.
As the guest owner only trusts CEKs for specific PSP firmware versions, she will dismiss the proposed CEK from the malicious cloud provider.
This prevents the deployment of SEV-protected guest VMs on SEV platforms with known security issues.

Since the \emph{Fake SEV} attack presented in Section \ref{subsec:fake_sev_attack} relies on a valid, extracted CEK, which can only come from a vulnerable and therefore revoked firmware, the new design effectively prevents this attack.

In a similar fashion, the \emph{Debug Override} attack presented in Section \ref{subsec:debug_override_attack} relies on the ability to alter the SEV firmware in order to patch and abuse its debug functionality.
This is only possible with a vulnerable, i.e., revoked, firmware, which will be dismissed by the client.
\\

\noindent\textbf{\textit{Migration}}\quad
Leveraging the enhanced SEV design, the guest owner successfully deployed her virtual machine on a trusted SEV platform.
To ensure the virtual machine is not migrated to a platform using a vulnerable firmware version, the source SEV firmware enforces a version check on the CEK.
The attack discussed in Section \ref{subsec:migration_attack} requires the attacker to provide an extracted CEK.
In the original SEV design, the source SEV firmware validates the CEK solely on the fact whether it has a root of trust originating in an ARK certificate.
In our enhanced design, the SEV firmware also ensures that the CEK certificate is valid for the SEV firmware version specified in the guest policy.
To that end, the CEK certificate contains the PSP OS version as well as the SEV firmware version numbers, see Section \ref{subsec:design_changes}.
The SEV firmware can now verify that the version fields of the provided CEK are equal to, or higher than the version numbers specified in the guest policy.

While an attacker could still provide a valid, extracted CEK, she cannot provide a CEK with a valid version field assuming the specified PSP firmware versions do not contain known security issues.
Migration of an SEV-protected virtual machine to an SEV firmware version lower than specified in the guest policy will not be permitted.
This protects against the \emph{Migration Attack} presented in Section \ref{subsec:migration_attack}.\\

%% file: src/7_related_work.tex
\section{Related Work}
\label{sec:related_work}
Work such as \cite{cloudvisor2011, hsvm2011, flicker2008, hypercoffer2013, hyperwall2012, fidelius2018} show that protecting virtual machines from an untrusted hypervisor has been subject to extensive research.

In \cite{cloudvisor2011} Zhang et al. propose the use of a higher-privileged security monitor called \emph{CloudVisor} in order to protect virtual machines in the face of a compromised hypervisor.
In their attack model, they assume the cloud provider himself not to be malicious and therefore exclude physical attacks.
In their design, that is very similar to the design of AMD SEV, they aim to separate resource management from security protection in the virtualization layer.
Their prototype version of CloudVisor is implemented in only 5.5K lines of code and works with the Xen hypervisor.
Zhang et al. leverage a Trusted Platform Module (TPM) in two ways:
The integrity of the CloudVisor code is ensured by the TPM and Intel Trusted Execution Technology (TXT).
The second use of a TPM in CloudVisor is to provide confidential communication between the cloud customer and CloudVisor.
It roots in the Storage Root Key (SRK) of the TPM, that has the lifetime of the platform owner and is authenticated by the Endorsement Key (EK) of the TPM.
The EK keypair of TPMs is unique per chip, thus binding the TPM's identity.
Its private key must therefore never leave the TPM.
The manufacturer of the TPM provides certificates of the EK public key through a certificate authority (CA).
Given that the specific identities of TPMs used for CloudVisor are guaranteed by a third party, this enables the cloud customer to also authenticate intermediate keys signed by the EK.
These can then be used for confidential communication.
With AMD SEV, the CEK serves a similar purpose as the EK of a TPM: It binds the identity of an SEV-capable CPU using chip-unique one-time-programmable fuses.

While the remote attestation mechanism of SEV has previously not been subject to research, researchers presented several attacks against the runtime protection of SEV.

In 2017 Hetzelt and Buhren presented the first security analysis of the SEV Technology \cite{Hetzelt:2017}.
They proposed three attacks on an SEV-enabled system, two of which rely on the lack of protection of the \textit{Virtual Machine Control Block} (VMCB) and registers. 
These are mitigated in case the \emph{encrypted state} extension to SEV is enabled, see Section \ref{subsec:sev}.
The third attack leverages the missing integrity protection of the guest memory, allowing the hypervisor to conduct a replay attack.
To that end, the nested pagetable was altered to enforce a pagefault on every guest page that was executed.
It was shown that the sequence of pagefaults is enough to determine the location of a password in the target VM's memory.
The page containing the password was then replayed during an attacker initiated SSH login.
Using this primitive, the authors were able to gain root access to a target system protected by SEV.

Morbitzer et al. leveraged the hypervisor's control over the guest resources, namely the nested pagetables, to mount an attack against an SEV-protected VM~\cite{DBLP:journals/corr/abs-1805-09604}.
In a similar fashion as ~\cite{Hetzelt:2017} they used page tracking to identify a guest page which is served via a server running in the VM.
As the memory is encrypted with a guest-specific key, the server will copy the data to an unencrypted page, e.g., the buffer of a virtual network card.
Once this page is located, they manipulate the nested pagetable to change the mapping between guest physical pages to host physical pages.
The new mapping will point to confidential data of the VM.
Instead of copying the intended data, the server will instead copy the confidential data into the unencrypted buffer.
This buffer is then readable by the attacker.
Using this method, the authors were able to fully decrypt a VM with 2GB of memory.

In 2018, Israeli research firm CTS Labs published a whitepaper called \emph{AMDFlaws} claiming to have found multiple critical vulnerabilities in AMD processors allowing arbitrary code execution on the PSP~\cite{CTSLABS2018}.
In the whitepaper, the researchers publish conceptual information about vulnerabilities in both Epyc and Ryzen PSPs as well as alleged manufacturer backdoors in both firmware and hardware of AMD chipsets supporting Ryzen CPUs. 
In 2019, two members of CTS Labs presented insights into three vulnerabilities of the AMDFlaws publication~\cite{CTSLABS2019}. 
In the presentation, the researchers illuminate details about the firmware of the PSP and the cryptographic checks in use. 
Nonetheless, the researchers did not discuss the consequences of a compromised PSP to the SEV technology.

%% file: src/8_conclusion.tex
\section{Conclusion}
\label{sec:conclusion}
In this paper, we analyzed the firmware components that implement the SEV API.
We identified security issues in the secure boot mechanism of the PSP that hosts the SEV firmware.
This allowed us to provide a patched version of the SEV firmware which gives us arbitrary read and write access to the PSP's memory.
We used this firmware to extract the \textit{Chip Endorsement Key} (CEK) of three different AMD Epyc processors.
We proposed two attacks against SEV-protected virtual machines using the extracted CEK as well as an attack based on a patched SEV firmware.
While the patched firmware allowed us to extract encrypted memory in plaintext, the extracted CEK allows an attacker to impersonate the presence of SEV altogether.
Even if the targeted virtual machine is not executed on a compromised SEV platform, the migration attack allows an attacker to acquire the cryptographic keys that are used to encrypt the virtual machines during migration.

The severity of the proposed attacks is amplified due to the missing rollback prevention as well as the infinite lifetime of the CEK.
We showed that an attacker can always roll back to a vulnerable PSP firmware to extract the CEK.
Even if the PSP firmware is upgraded to a newer version, this extracted CEK is still valid for the corresponding CPU. 

In the current design of the SEV technology, it is impossible for a cloud customer to verify the integrity of the remote platform given the fact that a vulnerable firmware version exists.
We conclude that the SEV technology on AMD Epyc systems of the Naples CPU series cannot protect virtual machines as the correct deployment cannot be guaranteed.
Given the lifetime of the CEK, it is not possible to provide purely software-based mitigations.

To overcome the issues of the current SEV technology, we proposed design changes to SEV that enable the cloud customer to enforce the use of a specific PSP firmware on the remote platform.
This ensures the trustworthiness of the SEV technology despite PSP firmware issues as it allows to issue software-based fixes for the PSP firmware.